\documentclass[twocolumn,superscriptaddress,nobalancelastpage,10pt,prr,letter]{revtex4-2}

\usepackage{amsmath}
\usepackage{amsthm}
\usepackage{amssymb}
\usepackage{amsfonts}
\usepackage{graphicx}
\usepackage{wasysym}
\usepackage{mathrsfs}
\usepackage{yfonts}
\usepackage{bbold}
\usepackage{verbatim}
\usepackage{subfigure}
\usepackage{color}
\usepackage{lipsum}
\usepackage{mathtools}
\usepackage{setspace}

\usepackage[normalem]{ulem}

\newcommand{\ket}[1]{| #1 \rangle}
\newcommand{\bra}[1]{\langle #1 |}

\newcommand{\opa}{\hat{a}}
\newcommand{\ocr}{\hat{a}^{\dag}}
\newcommand{\opEps}{\hat{E}^{(+)}}
\newcommand{\opEns}{\hat{E}^{(-)}}

\renewcommand{\r}{\mathbf{r}}
\renewcommand{\k}{\mathbf{k}}
\DeclareMathOperator{\sinc}{sinc}

\makeatletter
\newcommand*{\rom}[1]{\expandafter\@slowromancap\romannumeral #1@}
\makeatother

\begin{document}	
	
	\title{Phase-Subtractive Interference and Noise-Resistant Quantum Imaging with Two Undetected Photons}
	
	\author{Chandler Tarrant}
	\affiliation{Department of Physics, 145 Physical Sciences Bldg., Oklahoma State University, Stillwater, OK 74078, USA}
	
	\author{Mayukh Lahiri}
	\email{mlahiri@okstate.edu} \affiliation{Department of Physics, 145 Physical Sciences Bldg., Oklahoma State University, Stillwater, OK 74078, USA}

	\begin{abstract}
		We introduce and theoretically analyze a quantum interference phenomenon in which a two-photon interference pattern is created using four-photon quantum states generated by two independent sources and without detecting two of the photons. Contrary to the common perception, the interference pattern can be made fully independent of phases acquired by the photons detected to construct it. However, it still contains information about spatially dependent phases acquired by the two undetected photons. This phenomenon can be observed with both bosonic and fermionic particles. We show that the phenomenon can be applied to develop a quantum quantitative phase imaging technique that is immune to uncontrollable phase fluctuations in the interferometer and allows image acquisition without detecting the photons illuminating the object.
	\end{abstract}
	\maketitle

    \section{Introduction}\label{sec:intro}
	The principle of quantum superposition, when applied to two-particle systems, yields richer phenomena than can be observed in single-particle systems. One ``mind-boggling'' example \cite{greenberger1993multiparticle} is interference by \emph{path identity} of undetected photons \cite{Lahiri2022PI_RMP}, which was first reported by Zou, Wang and Mandel (ZWM) in the early 1990s \cite{zou1991induced,wang1991induced}. ZWM created a superposition of the origin of a photon pair and then controlled the interference of one of the photons by path identity of its partner photon. A counter-intuitive fact is that the resulting single-photon interference pattern can be manipulated by interacting with the partner photon which is never detected. Conventional interference by path identity relies on sources that emit coherently, i.e., sources that are not independent. We show that if independent quantum sources are used, an even more counter-intuitive and unusual interference phenomenon emerges. This phenomenon is a manifestation of the four-particle superposition principle and promises a significant advancement in the field of quantum imaging. 
	\par
	We consider a four-photon state generated by two independent quantum sources and show that two-photon interference patterns with unique properties can be created by path identity of two undetected photons. In standard two-particle interference experiments \cite{horne1989two}, phase differences associated with the two detected particles get added with the same sign when acquired in the way shown in Fig.~\ref{fig:2mode}a (Eq.~\eqref{2-ph-pattern} below). All reported two-particle interference effects obtained by path identity of undetected photons also display the same property \cite{lahiri2018many,qian2023multiphoton}. In contrast, the phases of the two detected particles get added with opposite signs in the interference phenomenon reported by us (Fig.~\ref{fig:2mode}b). We show by considering multi-mode photonic states that this fact can be used to develop a highly phase-stable interferometer, which produces interferograms that do not depend on the tunable interferometric phase but retain the information of any spatially dependent phase introduced to the undetected photons.
	\begin{figure}[htbp]
	\includegraphics[width=\linewidth]{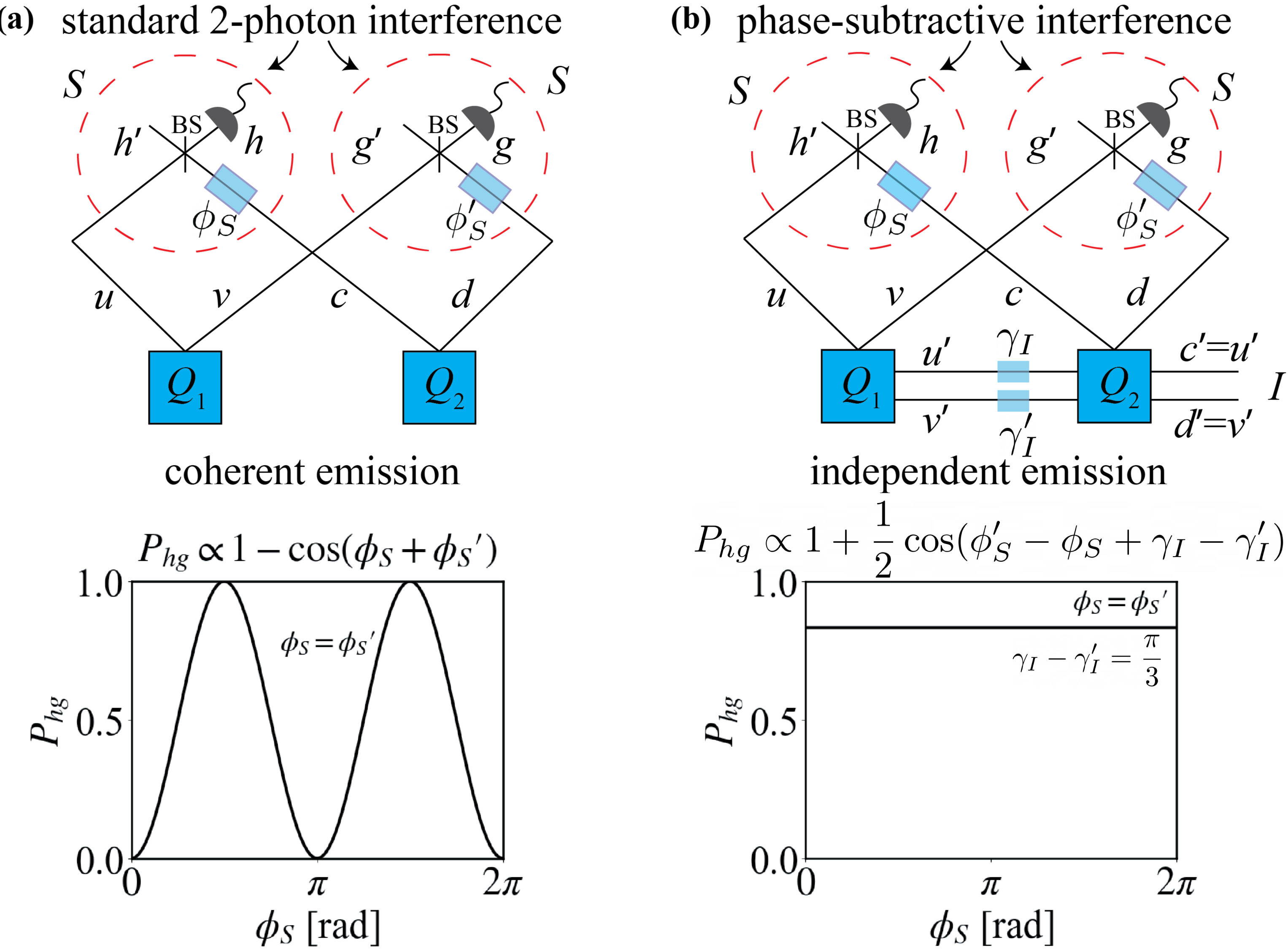}
		\caption{\textbf{(a)} Standard two-particle interference. Sources $Q_1$ and $Q_2$ coherently emit a pair of identical particles ($S$). The coincidence counting rate ($P_{hg}$) at beamsplitter (BS) outputs $g$ and $h$ varies sinusoidally with phase $\phi_S+\phi_S'$. \textbf{(b)} Phase-subtractive interference by path identity (PSIPI) of two particles. $Q_1$ and $Q_2$ are independent sources emitting two pairs of particles. A pair is made of particles $S$ and $I$. Two-particle interference at $g$ and $h$ is observed by detecting $S$-particles when path identity is employed using $I$-particles, i.e., when paths $u'$ and $v'$ are made identical with paths $c'$ and $d'$. The coincidence counting rate varies sinusoidally with $\phi_S-\phi_S'$ and also contains information of phases $\gamma_I$ and $\gamma_I'$ acquired by undetected particles. The interference pattern is independent of phases acquired by detected particles when $\phi_S=\phi_S'$.}
		\label{fig:2mode}
	\end{figure}
	\par
	Conventional interference by path identity has led to the development of a unique imaging technique, namely quantum imaging with undetected photons (QIUP) \cite{lemos2014quantum,lahiri2015theory,viswanathan2021position,kviatkovsky2022mid,lemos2022quantum}. This technique allows one to image an object without detecting photons that interacted with it. Consequently, QIUP allows one to determine optical properties of an object in spectral ranges for which adequate detectors are not available. However, QIUP is vulnerable to uncontrollable random fluctuations of the interferometric phase (phase noise) that arise due to the instability of the interferometer. Therefore, the photon acquisition time in QIUP cannot be long. Furthermore, if the phase noise destroys the interference pattern, QIUP becomes inapplicable. Recently, a noise-resistant interferometric phase imaging (NRIPI) technique has been introduced \cite{szuniewicz2023noise,thekkadath2023intensity}, which is inspired by interference of light from independent quantum sources \cite{pfleegor1967interference,Mandel1983,ou1997parametric,chrapkiewicz2016hologram,ou2007multi}. However, NRIPI must detect photons that interact with the object and is, therefore, inapplicable to spectral ranges where adequate detectors are not available. Here, we show that the interference phenomenon reported by us can be applied to develop a phase imaging technique which is immune to phase noise, allows arbitrarily long photon acquisition time, and can acquire images at wavelengths for which no detectors are available. Our imaging technique is fundamentally different from conventional \cite{pittman1995optical,gatti2008quantumimaging,chan2009two,aspden2013epr} and interaction-free \cite{zhang2019interaction} ghost imaging, which are solely two-photon phenomena.
    \par
    The paper is organized as follows: In Sec.~\ref{sec:two-part-interf}, we recall some basic features of standard two particle interference. In Sec.~\ref{sec:principle}, we provide a basic analysis of the novel interference effect which we call ``phase-subtractive interference by path identity" (PSIPI). Section \ref{sec:ph-stable} then gives an overview of how PSIPI can be used to make a phase-stable quantum imaging system. In Sec.~\ref{sec:multimode}, we generalize the basic analysis presented in the previous sections to the multimode scenario. Then, in Sec.~\ref{sec:Imaging}, we provide a numerical demonstration of quantitative phase imaging enabled by the principle of PSIPI. Section \ref{discuss} provides a brief discussion, followed by Sec.~\ref{summary} where we summarize our results and mention the outlook of our work.

\section{Standard Two-Particle Interference}\label{sec:two-part-interf}
We first recollect some basic features of standard two-particle interference (Fig.~\ref{fig:2mode}a). We consider two coherently and non-simultaneously emitting sources, $Q_1$ and $Q_2$, each of which can generate a pair of identical particles ($S$). The particle can be a boson or a fermion. Suppose now that $Q_1$ can emit the particle-pair into paths (modes) $u$ and $v$, and $Q_2$ can emit them into paths $c$ and $d$. Paths $u$ and $c$ are combined by a beamsplitter with outputs $h$ and $h'$. The tunable phase difference between paths $u$ and $c$ is denoted by $\phi_S$. Likewise, paths $v$ and $d$ are superposed with phase difference $\phi_S'$, and the corresponding beamsplitter outputs are $g$ and $g'$. Detectors are placed at outputs $h$ and $g$.
The probability of joint detection of two $S$-particles at $g$ and $h$ is given by (Appendix~\ref{sec:StandardInterference})
\begin{align}\label{2-ph-pattern}
P_{hg}\propto 1-\cos(\phi_S+\phi_S'),
\end{align}
which represents a typical two-particle interference pattern. We note that phases $\phi_S$ and $\phi_S'$ got added with the same sign. Such phases also get added in the same manner in \emph{conventional two-particle interference by path identity} \cite{lahiri2018many,qian2023multiphoton}.

\section{Basic Analysis of Phase-Subtractive Interference by Path Identity}\label{sec:principle}

\subsection{Description of the Interferometer}\label{subsec:setup}
We now examine the interferometer depicted by Fig.~\ref{fig:2mode}b. The interferometer contains two \emph{independent} sources, $Q_1$ and $Q_2$. We illustrate the setup using photonic states created by spontaneous parametric down conversion (SPDC) in nonlinear crystals (see, for example, Ref.~\cite{walborn2010spatial}; see also Appendix~\ref{sec:SPDC}). That is, $Q_1$ and $Q_2$  can be considered as two nonlinear crystals pumped by two \emph{mutually incoherent} laser beams. 
\par
SPDC is a quantum $\chi^{(2)}-$nonlinear process, i.e., at the basic level, the process of SPDC converts a pump photon into a photon pair. We assume that the two photons forming a pair can, in general, have different wavelengths and call them signal ($S$) and idler ($I$) following standard terminology. Four-photon states created by SPDC are essentially results of simultaneous emissions of two such photon pairs. 
\par
In order to understand the fundamental features of the phase-subtractive interference by path identity, it is instructive to consider a fully correlated two-mode scenario. In this case, each photon can be emitted into two possible paths, where each path corresponds to a distinct mode, for example, a momentum mode of the photon. (Multi-mode scenario is considered in Sec.~\ref{sec:multimode} below.) Suppose that $Q_1$ can emit a pair of photons into a pair of paths $(u,u')$ or $(v,v')$, where paths $u$ and $v$ can be occupied by the signal photon and paths $u'$ and $v'$ can be occupied by the idler photon (Fig.~\ref{fig:2mode}b). Likewise, $Q_2$ can emit a photon pair into a pair of paths $(c,c')$ or $(d,d')$. Consequently, paths $u$, $v$, $c$, and $d$ can only be occupied by signal photons, whereas idler photons can only be in paths $u'$, $v'$, $c'$, and $d'$. 
\par
Paths $u$ and $c$ are combined by a beamsplitter with outputs $h$ and $h'$. The tunable phase difference between paths $u$ and $c$ is denoted by $\phi_S$. Likewise, paths $v$ and $d$ are superposed with phase difference $\phi_S'$, and the corresponding beamsplitter outputs are $g$ and $g'$. The two beamsplitters can be spatially separated (Fig.~\ref{fig:2mode}b), or a single beamsplitter of finite size can be used for combining both pairs of paths (Fig.~\ref{fig:RandPhase}a). We are interested in the probability of coincidence detection of two signal photons at $g$ and $h$ (Figs.~\ref{fig:2mode}b~and~\ref{fig:RandPhase}a). Note that output pairs $(g,h')$, $(g',h)$, and $(g',h')$ can be equivalently considered. 
\par
As shown in Fig.~\ref{fig:2mode}b, paths ($u'$ and $v'$) of idler photons from $Q_1$ are sent through $Q_2$ and aligned with paths ($c'$ and $d'$, respectively) of idler photons originating from $Q_2$. That is, paths of idler photons from the two sources are made identical. Such an alignment is called path identity \cite{Lahiri2022PI_RMP}. We show below that due to this alignment a unique type of interference effect is observed in the coincidence detection of signal photons.
We emphasize that \emph{idler photons are not detected, and no further postselection is considered.}

\subsection{States Generated by the Two Sources before Path Identity}\label{subsec:state-no-PI}
\begin{figure*}
\includegraphics[width=\linewidth]{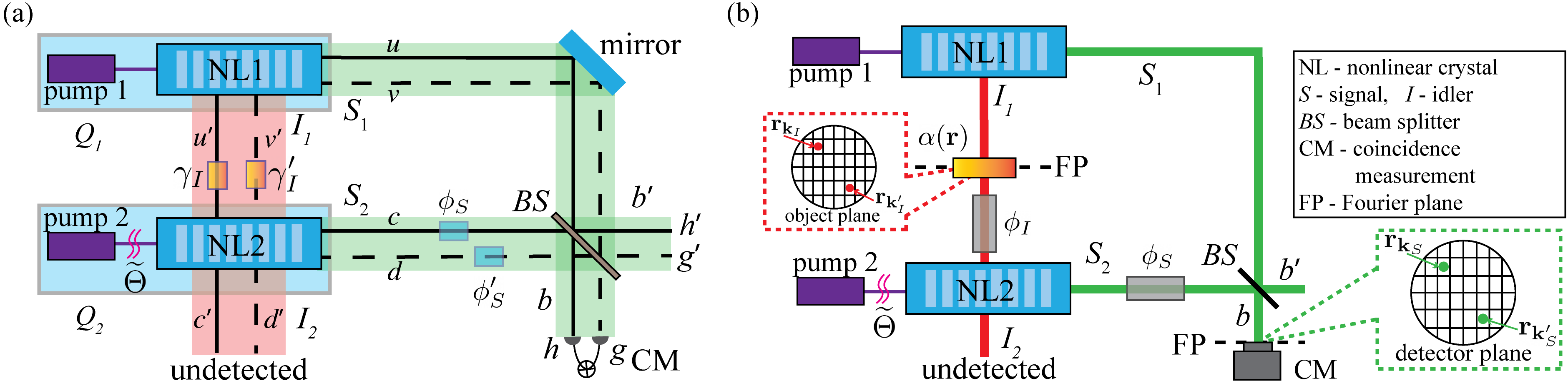}
\caption{\textbf{(a)} Principle of implementing PSIPI using photonic states. Two nonlinear crystals (NL1, NL2) are pumped with mutually incoherent laser beams. $\widetilde{\Theta}$ represents the phase difference between pump fields. Paths $u$ and $v$ ($c$ and $d$) taken by two signal photons generated by NL1 (NL2) lie within a highly collimated beam $S_1$ ($S_2$). A suitable lens system is used (not shown) to set $\phi_S=\phi_S'$. Paths $u'$ and $v'$, which are made identical with $c'$ and $d'$, lie within a highly collimated idler beam. Phases $\gamma_I$ and $\gamma_I'$ can only be different when a spatially dependent phase is introduced in the idler beam using a phase object. Signal beams, $S_1$ and $S_2$, are superposed by a beamsplitter (BS) and coincidence counts are measured at $g$ and $h$ in an output of BS. The coincidence counting rate gives a phase-subtractive interference pattern and it contains information of the phase object [Eq.~(\ref{img-2-mode})]. \textbf{(b)} Multimode case and proposed scheme for noise-resistant phase imaging with undetected photons (for an alternative setup, see Appendix \ref{sec:HRWZ}). It is a multimode version of the case considered in (a). The phase object and the camera are placed on the Fourier plane (FP) of the sources NL1 and NL2. Coincidence counts are measured at pairs of points $(\r_{\k_S},\r_{\k_S'})$ at an output of BS.} \label{fig:RandPhase}
\end{figure*}

Let us first consider the state generated by source $Q_1$ alone. As already mentioned, this source can emit a signal-idler pair in modes $(u,u')$ or $(v,v')$. We denote the corresponding states by $\ket{S_u,I_{u'}}_1$ and $\ket{S_v,I_{v'}}_1$, respectively. It follows from the standard theory of SPDC that the quantum state created by source $Q_1$ in the fully correlated two-mode scenario (Fig.~\ref{fig:2mode}b and \ref{fig:RandPhase}a) is given by (Appendix~\ref{sec:SPDC-2mode})
\begin{align}\label{2mode-state-1}
    \ket{\psi''}_1 &=\ket{\text{vac}} + gV_p \left(\ket{S_u,I_{u'}}_1+\ket{S_v,I_{v'}}_1 \right)\nonumber\\
    &+g^2V_p^2\Big(\,\ket{2S_u,2I_{u'}}_1+ \ket{2S_v,2I_{v'}}_1 \nonumber \\ & \qquad \quad \quad + \ket{S_u,I_{u'},S_v,I_{v'}}_1 \Big)+\cdots, 
\end{align}
where $|gV_p|\ll 1$ with $V_p$ representing the pump field at the crystal and $g$ being proportional to the gain of the crystal. In Eq.~\eqref{2mode-state-1}, the kets multiplied with $gV_p$ and $g^2V_p^2$ represent two-photon and four-photon states, respectively.
\par
We can express the quantum state given by Eq.~\eqref{2mode-state-1} as $\ket{\psi''}_1=\hat{U}_1 \ket{\text{vac}}$, where 
\begin{align}
&\hat{U}_1 = \hat{\openone} + gV_p \big\{\opa^\dagger_{S_1}(u)\opa^\dagger_{I1}(u') +\opa^\dagger_{S_1}(v)\opa^\dagger_{I1}(v') + \text{H.c.}\big\}\nonumber\\
&+\frac{g^2V_p^2}{2} \big\{ \big[ \opa^\dagger_{S_1}(u)\opa^\dagger_{I1}(u') \big]^2 +\big[\opa^\dagger_{S_1}(v)\opa^\dagger_{I_1}(v') \big]^2 \nonumber\\
& +2\opa^\dagger_{S_1}(u)\opa^\dagger_{I_1}(u') \opa^\dagger_{S_1}(v)\opa^\dagger_{I_1}(v') + \text{H.c.} \big\}+\cdots, \label{2mode-operator:a}
\end{align}
with $\opa_{S_1}^\dagger(u)$ representing the creation operator for a signal photon created by source $Q_1$ in mode $u$, etc., H.c. representing the Hermitian conjugate, and $\ket{\text{vac}}$ being the vacuum state.
\par
We now consider the fact that sources $Q_1$ and $Q_2$ are mutually independent. Such a situation arises, for example, when the two nonlinear crystals are pumped with mutually incoherent light beams (Fig.~\ref{fig:RandPhase}a). This situation can be analytically treated by introducing a random (stochastic) phase difference, $\widetilde{\Theta}$, between the pump fields at the two nonlinear crystals obeying
\begin{align}\label{rand-phase}
\langle \exp[i(\widetilde{\Theta}+\phi)] \rangle =\langle \cos (\widetilde{\Theta}+\phi) \rangle=\langle \sin (\widetilde{\Theta}+\phi) \rangle =0,
\end{align}
where $\phi$ is an arbitrary deterministic phase and the angular brackets represent averaged value. We note here that one can alternatively treat the problem using mixed states, which does not require introducing a random phase. However, the random phase-based approach is simpler and allows one to immediately apply the existing methodology \cite{Mandel1983,liu2009investigation,lahiri2019nonclassicality} that is based on pure states. Because of this, we adopt it in the main text, while we present the mixed-state based approach in Appendix~\ref{sec:DMT}. We emphasize that the two approaches are equivalent and yield the same result.
\par
The quantum state generated individually by $Q_2$ can now be expressed as $\ket{\psi''}_2=\hat{U}_2 \ket{\text{vac}}$ with 
\begin{align}
&\hat{U}_2 = \hat{\openone} + gV_p e^{i\widetilde{\Theta}} \big\{\opa^\dagger_{S_2}(c)\opa^\dagger_{I1}(c') +\opa^\dagger_{S_2}(d)\opa^\dagger_{I1}(d') \nonumber\\
&+ \text{H.c.}\big\} +\frac{g^2V_p^2}{2} e^{2i\widetilde{\Theta}} \big\{ \big[ \opa^\dagger_{S_2}(c)\opa^\dagger_{I1}(c') \big]^2 \nonumber\\
& +\big[\opa^\dagger_{S_2}(d)\opa^\dagger_{I1}(d') \big]^2 +2\opa^\dagger_{S_2}(c)\opa^\dagger_{I_2}(c') \opa^\dagger_{S_2}(d)\opa^\dagger_{I_2}(d') \nonumber\\
& + \text{H.c.} \big\}+\cdots, \label{2mode-operator:b}
\end{align}
where $\widetilde{\Theta}$ obeys Eq.~\eqref{rand-phase}, and we have assumed for simplicity that the pump beams at the two sources have equal intensity.
\par
When the two sources ($Q_1$ and $Q_2$) are pumped together, the quantum state jointly created by the two sources is given by 
\begin{align}\label{state-joint-1}
\ket{\psi'} = \hat{U}_2\hat{U}_1\ket{\text{vac}}.
\end{align}
where it can be readily checked that $\hat{U}_1$ and $\hat{U}_2$ commute. It now follows from Eqs.~\eqref{2mode-operator:a},~\eqref{2mode-operator:b}, and \eqref{state-joint-1} that (dropping a normalization constant)
\begin{align}\label{full-2mode-state-supp}
&\ket{\psi'} = \ket{\text{vac}} +gV_p \Big\{\ket{S_u,I_{u'}}_1+\ket{S_v,I_{v'}}_1\nonumber\\
&+e^{i\widetilde{\Theta}}\big(\ket{S_c,I_{c'}}_2+\ket{S_d,I_{d'}}_2\big)\Big\} \nonumber\\
&+g^2V_p^2 \Big\{\ket{2S_u,2I_{u'}}_1 +\ket{2S_v,2I_{v'}}_1+\ket{S_u,I_{u'},S_v,I_{v'}}_1\nonumber\\
&+e^{2i\widetilde{\Theta}}\Big(\ket{2S_c,2I_{c'}}_2+\ket{2S_d,2I_{d'}}_2 +\ket{S_c,I_{c'},S_d,I_{d'}}_2\Big)\nonumber\\
&+ e^{i\widetilde{\Theta}}\big(\ket{S_u,I_{u'}}_1\ket{S_c,I_{c'}}_2+\ket{S_u,I_{u'}}_1\ket{S_d,I_{d'}}_2\nonumber\\
&+\ket{S_v,I_{v'}}_1\ket{S_c,I_{c'}}_2+\ket{S_v,I_{v'}}_1\ket{S_d,I_{d'}}_2\big)\Big\},
\end{align}
where we have kept up to four-photon terms because we will find soon that the first non-vanishing contribution to the probability of coincidence detection comes from the four-photon terms of this perturbation series. Equation~\eqref{full-2mode-state-supp} represents the state created by two sources before path identity has been applied.
\par
The detection of photons will be treated formally in Sec.~\ref{subsec:detection} below. However, noticing some aspects of the detection at this stage helps simplify the analysis significantly as well as gain physical insights into the phenomenon.
\par
As shown in Fig.~\ref{fig:2mode}b, modes $u$ and $c$ are combined by a beamsplitter with outputs $h$ and $h'$. Likewise, modes $v$ and $d$ are superposed and the corresponding beamsplitter outputs are $g$ and $g'$. We are interested in the probability of coincidence detection of two signal photons at $g$ and $h$. (Note that output pairs $(g,h')$, $(g',h)$, and $(g',h')$ can be equivalently considered.) It is evident from Eq.~\eqref{full-2mode-state-supp} that two signal ($S$) photons are simultaneously present only in four-photon and higher order terms. Therefore, the first nonvanishing contribution to the coincidence counting rate comes from the four-photon terms. Consequently, all two-photon terms and $\ket{\text{vac}}$ in Eq.~\eqref{full-2mode-state-supp} can be dropped. We also drop the higher-order terms because their contribution is insignificant compared to four-photon terms \textemdash a standard procedure in perturbative treatment.
\par
Furthermore, not all four-photon terms in Eq.~\eqref{full-2mode-state-supp} contribute to the coincidence counting rate at $g$ and $h$ (and equivalently at output pairs $(g,h')$, $(g',h)$, and $(g',h')$). For example, $\ket{2S_u,2I_{u'}}_1$ contains two signal photons in mode $u$. However, it is evident from Fig.~\ref{fig:2mode}b that no photon in mode $u$ can reach $g$ (or $g'$). Therefore, the term $\ket{2S_u,2I_{u'}}_1$ does not contribute to the coincidence counting rate. Likewise, the terms $\ket{2S_v,2I_{v'}}_1$, $\ket{2S_c,2I_{c'}}_2$, $\ket{2S_d,2I_{d'}}_2$, $\ket{S_u,I_{u'}}_1\ket{S_c,I_{c'}}_2$, and $\ket{S_v,I_{v'}}_1\ket{S_d,I_{d'}}_2$ also do not contribute. As a result, the quantum state which contributes to two-photon coincidence counting at $g$ and $h$ is given by
\begin{align}\label{2-mode-state}
\ket{\tilde{\psi}} &= \ket{S_u,I_{u'},S_v,I_{v'}}_1 +e^{2i\widetilde{\Theta}}\ket{S_c,I_{c'},S_d,I_{d'}}_2 \nonumber\\
&+e^{i\widetilde{\Theta}} \big(\ket{S_u,I_{u'}}_1\ket{S_d,I_{d'}}_2+\ket{S_v,I_{v'}}_1 \ket{S_c,I_{c'}}_2\big),
\end{align}
where we have dropped a normalization coefficient. We will use Eq.~\eqref{2-mode-state} in our analysis. We, however, stress that one can instead work with Eq.~\eqref{full-2mode-state-supp} and obtain the same result.
\par
Equation~\eqref{2-mode-state} implies that there are three distinct ways in which a coincidence detection at $g$ and $h$ (and equivalently at output pairs $(g,h')$, $(g',h)$, and $(g',h')$) can occur (Fig.~\ref{fig:2mode}b): 
\begin{enumerate}
\item
A double-pair production occurs in $Q_1$ creating the state $\ket{S_u,I_{u'},S_v,I_{v'}}_1$. 
\item
A double-pair production occurs in $Q_2$ creating the state $\ket{S_c,I_{c'},S_d,I_{d'}}_2$. 
\item
Simultaneous single-pair productions occur at sources 1 and 2 with two options:
\begin{itemize}
\item[(a)] 
creating $\ket{S_u,I_{u'}}_1\ket{S_d,I_{d'}}_2$, and 
\item[(b)]
creating $\ket{S_v,I_{v'}}_1 \ket{S_c,I_{c'}}_2$.
\end{itemize}
\end{enumerate}
Equation~\eqref{2-mode-state} shows that the random phase $\widetilde{\Theta}$ does not appear between the kets corresponding to options 3(a) and 3(b) despite the two sources being independent. In fact, these two options can be made indistinguishable by applying path identity, leading to interference effects. Whereas, options $1$ and $2$ are both fully distinguishable from the rest of the options since phase differences between the corresponding kets always involve the random phase $\widetilde{\Theta}$. Therefore, options 1 and 2 cannot contribute to interference.

\subsection{States Generated by the Two sources after Path Identity}\label{subsec:state-PI}
We now apply path identity \cite{Lahiri2022PI_RMP}: paths of idler photons from $Q_1$ are sent through $Q_2$ and aligned with paths of idler photons originating from $Q_2$ (Fig.~\ref{fig:2mode}b), i.e., paths of idler photons from the two sources are made identical. Consequently, quantum fields (annihilation operators) corresponding to idler ($I$) photons generated by $Q_1$ and $Q_2$ become related by
\begin{align}\label{PI-field}
\opa_{I_2}(c')=e^{i\gamma_I}\opa_{I_1}(u') \quad \text{and} \quad \opa_{I_2}(d')=e^{i\gamma_I'}\opa_{I_1}(v'),
\end{align}
where $\gamma_I$ and $\gamma_I'$ are phases introduced to paths $u'$ and $v'$, respectively, between the two sources. Since $\opa^\dagger_{I_1}(u') \ket{\text{vac}}\equiv\ket{I_{u'}}_1$, etc., the path identity implies the following simultaneous relations involving kets: 
\begin{align}\label{PI-state}
\ket{I_{c'}}_2= e^{-i\gamma_I}\ket{I_{u'}}_1 \quad \text{and} \quad \ket{I_{d'}}_2= e^{-i\gamma_I'} \ket{I_{v'}}_1.
\end{align}
Using Eq.~\eqref{PI-state}, we find that Eq.~(\ref{2-mode-state}) reduces to
\begin{align}\label{2-mode-state-PI}
&\ket{\psi}= \ket{I_{u'},I_{v'}}_1 \otimes \big\{\ket{S_u,S_v}_1 +e^{i(2\widetilde{\Theta}-\gamma_I-\gamma_I')}\ket{S_c,S_d}_2 \nonumber\\
&+e^{i(\widetilde{\Theta}-\gamma_I')} \big[ \ket{S_u}_1\ket{S_d}_2+e^{-i(\gamma_I-\gamma_I')}\ket{S_v}_1\ket{S_c}_2 \big] \big\},
\end{align}
where we have dropped a normalization coefficient. Equation~(\ref{2-mode-state-PI}) represents the effective quantum state generated by two independent sources before beamsplitters in Fig.~\ref{fig:2mode}b and Fig.~\ref{fig:RandPhase}a.
\par
It is evident from Eq.~\eqref{2-mode-state-PI} that the idler states factor out\textemdash a trait of path identity. Furthermore, the quantum state contains a linear superposition of signal states $\ket{S_u}_1\ket{S_d}_2$ and $\ket{S_v}_1\ket{S_c}_2$ with a phase difference $\gamma_I-\gamma_I'$. This fact indicates that two-photon interference will be observed if the signal photons are detected as shown in Figs.~\ref{fig:2mode}b and \ref{fig:RandPhase}a, and the interference pattern will contain information of the phases acquired by idler photons\textemdash a result we explicitly show in the next section.
    
\subsection{Detection and Coincidence Counting Rates}\label{subsec:detection}
Signal modes $u$ and $c$ are combined by a balanced beamsplitter with outputs $h$ and $h'$ (Fig.~\ref{fig:2mode}b). The tunable phase difference between paths corresponding to $u$ and $c$ is denoted by $\phi_S$. Likewise, modes $v$ and $d$ are superposed with phase difference $\phi_S'$, and the corresponding beamsplitter outputs are $g$ and $g'$. Therefore, the positive frequency parts of the quantized electric fields corresponding to signal photons at the detectors placed at beamsplitter-outputs $h$ and $g$ are given by
\begin{subequations}\label{det-field}
\begin{align}
&\opa_{S}(h) \propto \opa_{S_1} (u)+ie^{i\phi_S}\opa_{S_2}(c), \label{det-field:a} \\
&\opa_{S}(g) \propto \opa_{S_1}(v)+ie^{i\phi_S'}\opa_{S_2}(d). \label{det-field:b}
\end{align}
\end{subequations}
The probability of coincidence detection (coincidence counting rate) of two signal photons at $g$ and $h$ can now be determined using the standard formula \cite{MW}
\begin{align}\label{coin-form}
P_{hg} \propto \bra{\psi}\opa^\dagger_S(h) \opa^\dagger_S(g) \opa_S(g) \opa_S(h) \ket{\psi},
\end{align}
where $\ket{\psi}$ is given by Eq.~(\ref{2-mode-state-PI}). 
\par
Using Eqs.~(\ref{rand-phase}), (\ref{2-mode-state-PI}), \eqref{det-field}, and \eqref{coin-form}, we readily find that the coincidence counting rate at $g$ and $h$ is given by
\begin{align}\label{2-ph-PI}
P_{hg} \propto 1+\frac{1}{2}\cos(\phi_S'-\phi_S+\gamma_I-\gamma_I').
\end{align}
Equation (\ref{2-ph-PI}) represents a unique two-photon interference pattern. In Eq.~(\ref{2-ph-PI}), \emph{phases $\phi_S$ and $\phi_S'$ got added with opposite signs} in striking contrast to standard two-particle interference (see Eq.~\eqref{2-ph-pattern}) and conventional two-particle interference by path identity \cite{lahiri2018many,qian2023multiphoton}. Therefore, if $\phi_S=\phi_S'$, the interference pattern becomes independent of phases gained by detected photons (Fig.~\ref{fig:2mode}b, bottom). In an experiment, $\phi_S$ and $\phi_S'$ can be set practically equal to each other (see Sec.~\ref{sec:ph-stable} below). Furthermore, if $\gamma_I \neq \gamma_I'$, the interference pattern contains information of phases introduced to idler photons that were never detected to construct the interference pattern. We call such an interference \emph{phase-subtractive interference by path identity} (PSIPI).
\par
The visibility of the interference pattern given by Eq.~(\ref{2-ph-PI}) is less than unity ($1/2$) because signal photons emitted individually by each source do not interfere and result in background. This is expected because the sources are independent: one can observe in Eq.~(\ref{2-mode-state-PI}) that the random phase difference, $\widetilde{\Theta}$, appears between kets $\ket{S_u,S_v}_1$ and $\ket{S_c,S_d}_2$ representing signal photons generated individually by $Q_1$ and $Q_2$, respectively. In contrast, the signal photons generated by the joint (simultaneous) emissions of $Q_1$ and $Q_2$ interfere; we observe in Eq.~(\ref{2-mode-state-PI}) that the random phase difference, $\widetilde{\Theta}$, does not appear between the corresponding states $\ket{S_u}_1\ket{S_d}_2$ and $\ket{S_v}_1\ket{S_c}_2$. In fact, the cosine term in Eq.~\eqref{2-ph-PI} arises solely from to the contribution of the superposition term $\ket{S_u}_1\ket{S_d}_2+e^{-i(\gamma_I-\gamma_I')}\ket{S_v}_1\ket{S_c}_2$ in Eq.~(\ref{2-mode-state-PI}).  It is possible to obtain interference fringes with unit visibility if the background due to the individual emissions is subtracted.
\par
In an experiment, lengths of interferometer arms need to be chosen appropriately such that interfering photons remain temporally indistinguishable as they arrive at the detectors. We recall that the cosine term in Eq.~(\ref{2-ph-PI}) arises solely due to the contribution from the state $\ket{I_{u'},I_{v'}}_1 \otimes (\ket{S_u}_1\ket{S_d}_2+e^{-i(\gamma_I-\gamma_I')}\ket{S_v}_1\ket{S_c}_2)$ in Eq.~(\ref{2-mode-state-PI}). This term, which contains a superposition of $\ket{S_u}_1\ket{S_d}_2$ and $\ket{S_v}_1\ket{S_c}_2$, is a result of simultaneous single-pair productions at two distinct sources that can happen in two alternative ways. Therefore, the lengths of interferometer arms need to be such that coincidence detections of $\ket{S_u}_1\ket{S_d}_2$ and $\ket{S_v}_1\ket{S_c}_2$ are indistinguishable when path identity is implemented.

\section{Principles of Phase Stability and Imaging}\label{sec:ph-stable}
We now show that PSIPI enables building a highly phase-stable interferometer applicable to imaging. Before considering a multi-mode scenario, we explain the principles using the two-mode scenario considered thus far.
\par
We first discuss how phases $\phi_S$ and $\phi_S'$ (Fig.~\ref{fig:2mode}b) can be set practically equal to each other. When an optical field travels through an interferometer, it acquires phases due to propagation through air (vacuum) and manipulation (e.g., reflection, refraction) by various optical elements (e.g., mirrors, lenses). We will call a phase acquired in this manner a \emph{propagation-phase}. In interferometry-based imaging techniques, propagation-phases are usually made spatially-independent by using highly collimated beams and appropriate lens systems \cite{Popescu-book,lemos2014quantum,kviatkovsky2020microscopy,szuniewicz2023noise}. Therefore, it is always possible to restrict paths $u$ and $v$ within a highly collimated beam (Fig.~\ref{fig:RandPhase}a) \textemdash for example, by using a lens system as in \cite{lemos2014quantum}\textemdash and set the propagation phases acquired along them practically equal. Likewise, propagation phases gained along paths $c$ and $d$ can also be made equal. Therefore, the phase difference, $\phi_S$, which arises due to optical path difference along $u$ and $c$ becomes practically equal to the phase difference, $\phi_S'$, that arises due to optical path difference along $v$ and $d$, i.e., 
\begin{align}\label{S-ph-eq}
\phi_S' -\phi_S =0. 
\end{align}
\par
We now consider a situation in which a phase object, which introduces a spatially dependent phase to an optical field passing through it, is inserted into the idler beam between the two sources. Let the phases introduced by the phase object to idler photons in paths $u'$ and $v'$ be $\alpha$ and $\alpha'$, respectively. In this case, the phase acquired by an idler photon has two parts: one is the propagation-phase ($\phi_I$ and $\phi_I'$), and the other is introduced by the phase object ($\alpha$ and $\alpha'$). That is, 
\begin{align}\label{I-tot-phase}
\gamma_I=\phi_I+\alpha, \quad \text{and} \quad \gamma_I' = \phi_I'+\alpha'. 
\end{align}
The propagation phases, $\phi_I$ and  $\phi_I'$, can be made practically equal to each other as discussed above. Consequently, it follows from Eq.~\eqref{I-tot-phase} that
\begin{align}\label{I-ph-eq}
\gamma_I-\gamma_I' = \alpha -\alpha'. 
\end{align}
\par
Using the relations given by Eqs.~\eqref{S-ph-eq} and \eqref{I-ph-eq}, we find that Eq.~(\ref{2-ph-PI}) reduces to
\begin{align}\label{img-2-mode}
P_{hg} \propto 1+\frac{1}{2}\cos(\alpha -\alpha').
\end{align}
We observe that the interference pattern represented by Eq.~(\ref{img-2-mode}) does not depend on spatially-independent propagation-phases (e.g., $\phi_S$, $\phi_I$) associated with signal, idler, and pump photons. Such phases are subject to random fluctuations due to instability of the interferometer. The interference pattern depends only on the phases introduced by the phase object ($\alpha$, $\alpha'$), which are fixed for a given object and do not fluctuate. Consequently, the interference pattern is immune to random phase fluctuations associated with instability of the interferometer while it retains the information of the object. Using established techniques, the spatially dependent phases introduced by the object can be retrieved from the interference pattern; that is, the phenomenon enables quantitative phase imaging. 
    
\section{Multimode Scenario}\label{sec:multimode}
We now consider the multimode scenario (Fig.~\ref{fig:RandPhase}b), in which there are an infinite number of momentum (wave vector) modes. We also no longer assume that the modes are perfectly correlated. The theoretical treatment is strictly similar to that used for the two mode case in Sec.~\ref{sec:principle}.
\par
In the multimode case, it follows from the theory of SPDC (Appendix~\ref{sec:SPDC}) that the state generated by an individual crystal (NL$j$ with $j=1,2$) is given by $\ket{\psi''}_j=\hat{U}_j \ket{\text{vac}}$, where
\begin{align}\label{gen-uhat}
    &\hat{U}_j = \hat{\openone}+ \eta_j \sum_{\k_S,\k_I}C^{(2)}(\k_S,\k_I)\opa^\dagger_{S_j}(\k_S)\opa^\dagger_{I_j}(\k_I) \nonumber\\
    &+ \eta_j^2\sum_{\substack{\k_S,\k_S'\\ \k_I,\k_I'}}C^{(4)}(\k_S,\k_S',\k_I,\k_I') \nonumber\\
    & \times \opa^\dagger_{S_j}(\k_S)\opa^\dagger_{S_j}(\k_S')\opa^\dagger_{I_j}(\k_I)\opa^\dagger_{I_j}(\k_I')+\text{H.c.}+\cdots,
\end{align}
with $\eta_1=1$ and $\eta_2=\exp(i\widetilde{\Theta})$. In Eq.~\eqref{gen-uhat}, $C^{(2)}(\k_S,\k_I)$ is the joint probability amplitude of photon-pair emission with signal and idler photons being in momentum-modes $\k_{S}$ and $\k_{S}$, respectively, and $C^{(4)}(\k_S,\k_S',\k_I,\k_I') = C^{(2)}(\k_S,\k_I)C^{(2)}(\k_S',\k_I')/2$. (An explicit form of $C^{(2)}(\k_S,\k_I)$ is given in Appendix~\ref{coin-mm-app}.) It can be readily checked that Eq.~(\ref{gen-uhat}) represents the multimode version of Eqs.~\eqref{2mode-operator:a} and \eqref{2mode-operator:b}.
\par
The quantum state jointly created by the two crystals is once again given by Eq.~\eqref{state-joint-1} with $\hat{U}_1$ and $\hat{U}_2$ given by Eq.~\eqref{gen-uhat}.
\par
Without any loss of generality, we choose the far-field configuration used in Refs.~\cite{lemos2014quantum,lahiri2015theory}. In this configuration, appropriate lens systems are used to (i) image crystal 1 onto crystal 2, and (ii) place the phase object and the camera (detector) on the Fourier plane of both sources (Fig.~\ref{fig:RandPhase}b). Consequently, distinct points on the phase object are impinged by idler photons with distinct momenta, and distinct points on the camera are impinged by signal photons with distinct momenta. We denote a point on the phase object (camera) corresponding to idler-momentum $\k_I$ (signal-momentum $\k_S$) by $\r_{\k_I}$ ($\r_{\k_S}$). In this case, the path-identity relation between idler kets produced at NL2 and NL1 becomes
\begin{align}\label{i-alignment-gen}
\ket{\k_I}_2=\exp[-i\{\phi_I+\alpha_I(\r_{\k_I}) \}] \, \ket{\k_I}_1,
\end{align}
where $\phi_I$ is a spatially independent phase acquired by the propagation of the idler photon between sources NL1 and NL2, and $\alpha(\r_{\k_I})$ is the spatially dependent phase introduced by the object.
\par
The quantum state created by the two sources after path identity is employed is determined by using Eqs.~\eqref{state-joint-1}, \eqref{gen-uhat}, and \eqref{i-alignment-gen}. The terms contributing to coincidence counting rates at an arbitrary pair of points ($\r_{\k_S},\r_{\k_S'}$) on the camera are found to be given by
\begin{align}\label{state-four-ZWM-Supp} 
&\ket{\psi^{(4)}}=\sum_{\substack{\k_S,\k_S' \\ \k_I,\k_I'}} \ket{\k_I,\k_I'}_1 \otimes \Big\{ C^{(4)}(\k_S,\k_S',\k_I,\k_I') \ket{\k_S,\k_S'}_1 \nonumber\\
&+ e^{i[2\widetilde{\Theta}-\gamma_I(\r_{\k_I})-\gamma_I(\r_{\k_I'})]} C^{(4)}(\k_S,\k_S',\k_I,\k_I')\ket{\k_S,\k_S'}_2 \nonumber \\
&+e^{i[\widetilde{\Theta}-\gamma_I(\r_{\k_I'})]} C^{(2)}(\k_S,\k_I)C^{(2)}(\k_S',\k_I')  \ket{\k_S}_1 \ket{\k_S'}_2  \Big\},
\end{align}                    
This state can be readily identified as the multimode generalization of Eq.~\eqref{2-mode-state-PI}. 
The three terms on the right-hand side of Eq.~(\ref{state-four-ZWM-Supp}) arise, respectively, due to emission from NL1 only, emission from NL2 only, and joint emissions from NL1 and NL2. 
\par
A general expression for the probability of coincidence detection at a pair of points $(\r_{\k_S},\r_{\k_S'})$ is given by 
\begin{align}\label{coin-2-supp}
&P_{jl}^{(2)}(\r_{\k_S},\r_{\k_S'}) \nonumber\\
&=\bra{\psi^{(4)}}\opEns_l(\r_{\k_S'})\opEns_j(\r_{\k_S})\opEps_j(\r_{\k_S'})\opEps_l(\r_{\k_S})\ket{\psi^{(4)}},
\end{align}
where $j=b,b'$ and $l=b,b'$ represent outputs of the beamsplitter $BS$ in Fig.~\ref{fig:RandPhase}b, $\opEps_j(\r_{\k_S'})$ represents the positive frequency part of the quantized electric field at output $j$, $\opEns=\{\opEps\}^\dagger$, and $\ket{\psi^{(4)}}$ is given by Eq.~(\ref{state-four-ZWM-Supp}). 
\begin{figure*}
\includegraphics*[width=\linewidth]{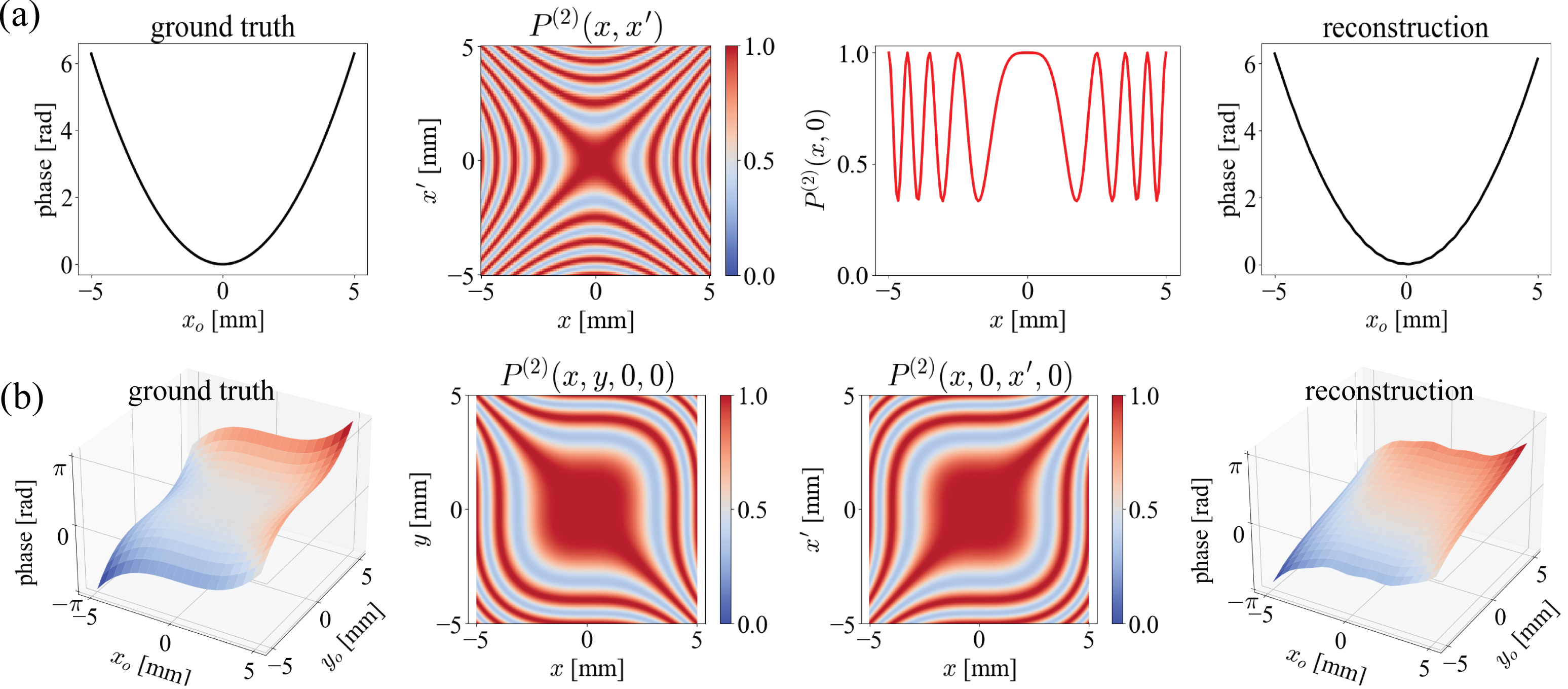}
\caption{Numerically simulated interferograms. Two-photon phase-stable interference patterns are created by detecting signal photons (810 nm) only, but they contain information of spatially dependent phase, $\alpha$, introduced to undetected idler photons (1550 nm). \textbf{(a)} Ground truth one-dimensional phase $\alpha(x_o)\propto x_o^2$ (leftmost) results in the coincidence counting rate (left-middle). Coincidence rates along the line $x'=0$ on the detector are also shown (right-middle). Phase image numerically-reconstructed from coincidence counting rate (rightmost). \textbf{(b)} Ground truth two-dimensional phase $\alpha(x_o,y_o)\propto x_o^3+y_o^3$ (leftmost) leads to four-dimensional coincidence map $P^{(2)}(x,y;x',y')$. The $xy$-plane cross-section (left-middle) and $xx'$-plane cross-section (right-middle) of the coincidence map are shown. Phase image numerically reconstructed from full four-dimensional coincidence counting rate (rightmost).}
\label{fig:Phase-Plots} 
\end{figure*}
\par
For our purpose, it is enough to choose both $\r_{\k_S}$ and $\r_{\k_S'}$ to be located at the same beamsplitter output, $b$, say (Fig.~\ref{fig:RandPhase}b). (Expressions for the other cases in given in Appendix~\ref{coin-mm-app}.) The positive frequency part of the quantized electric field operators at an arbitrary point in beamsplitter outputs $b$ can be written as
\begin{align}\label{b_output}
\opEps_b(\r_{\k_S})\propto \opa_{S_1}(\k_S)+ie^{i\phi_S}\opa_{S_2}(\k_S).
\end{align}
Using Eqs.~(\ref{state-four-ZWM-Supp}),~(\ref{coin-2-supp}), and (\ref{b_output}), we find that the the probability of coincidence detection (coincidence counting rate) is given by 
\begin{align}\label{coin-rate}
&P_{bb}^{(2)}(\r_{\k_S},\r_{\k_S'}) \propto \sum_{\k_I,\k_I'}\bigg[P^{(4)}_{1}(\k_S,\k_S',\k_I,\k_I')\nonumber\\
&+P^{(4)}_{2}(\k_S,\k_S',\k_I,\k_I') \nonumber\\
&+2P^{(4)}_{12}(\k_S,\k_S',\k_I,\k_I') \big\{1+\cos [\alpha(\r_{\k_I})-\alpha(\r_{\k_I'})]\big\}\bigg]
\end{align}
where $P^{(4)}_{1}$ and $P^{(4)}_{2}$ are individual contributions to the coincidence detection probability from NL1 and NL2, respectively, and the term  containing $P^{(4)}_{12}$ is due to the joint contribution from NL1 and NL2 (explicit forms are given in Appendix~\ref{coin-mm-app}). 
\par
Equation (\ref{coin-rate}) represents the phase-stable interference pattern produced by PSIPI in a general multimode scenario. It is the multimode version of Eq.~(\ref{img-2-mode}), where points $\r_{\k_S}$ and $\r_{\k_S'}$ correspond to $g$ and $h$.

\section{Illustration and Application to Imaging}\label{sec:Imaging}
We now illustrate the results and numerically demonstrate how a phase object can be reconstructed from the coincidence counting rate. For simplicity, we restrict ourselves to the case of maximum momentum correlation in which signal and idler photons are delta-correlated in momenta. For example, a highly collimated pump beam in phase-matched nonlinear crystals practically allows us to write $ C^{(2)}(\k_S,\k_I)\propto \delta_{\k_S,\k_0-\k_I}$ (see Appendix~\ref{sec:SPDC}, Eq.~\eqref{perf-MC}). It follows from the explicit forms of $P^{(4)}_{1}(\k_S,\k_S',\k_I,\k_I')$, $P^{(4)}_{2}(\k_S,\k_S',\k_I,\k_I')$, and $P^{(4)}_{12}(\k_S,\k_S',\k_I,\k_I')$ given in Appendix~\ref{coin-mm-app} that in this case Eq.~\eqref{coin-rate} reduces to
\begin{align}\label{perf-corr-coin-rate}
P_{bb}^{(2)}(\r_{\k_S},\r_{\k_S'})\propto 1+\frac{1}{2}\cos [\alpha(\r_{\k_S})-\alpha(\r_{\k_S'})].
\end{align}
\par
For illustration, we choose the signal wavelength to be $810$~nm and the idler wavelength to be $1550$~nm. We represent a point on the object ($\r_{\k_I}$) and a point on the camera ($\r_{\k_S}$) by $\r_{\k_I}\equiv (x_o,y_o)$ and $\r_{\k_S}\equiv (x,y)$. We consider a one-dimensional (1D) and a two-dimensional (2D) phase object represented by the following phase profiles, respectively:
\begin{itemize}
    \item[(a)]
    $\alpha(x_o)\propto x_o^2$,
    \item[(b)]
    $\alpha(x_o,y_o)\propto x_o^3+y_o^3$.
\end{itemize}
For the 1D phase object, the coincidence counting rate depends on two coordinates $(x,x')$, i.e., $P^{(2)}(\r_{\k_S},\r_{\k_S'})\equiv P^{(2)}(x,x')$. Figure~\ref{fig:Phase-Plots}a (left-middle) illustrates the corresponding two-photon interference pattern. For the 2D phase object, the coincidence rate depends on four coordinates $(x,y,x',y')$, i.e., $P^{(2)}(\r_{\k_S},\r_{\k_S'})\equiv P^{(2)}(x,y,x',y')$. Figure \ref{fig:Phase-Plots}b shows two cross-sections of the coincidence map: $xy$-plane (left-middle) and $xx'$-plane (right-middle), each displaying interference patterns.
\par
To numerically demonstrate quantitative phase imaging \cite{Popescu-book}, we have reconstructed the one-dimensional (Fig.~\ref{fig:Phase-Plots}a, rightmost) and two-dimensional (Fig.~\ref{fig:Phase-Plots}b, rightmost) phase objects. The method employs standard phase retrieval algorithms \cite{herraez2002fast,mertz2019introduction}. The application of such algorithms to coincidence maps has been demonstrated in Refs. \cite{szuniewicz2023noise,thekkadath2023intensity}.
    
\section{Discussion}\label{discuss}
Phase-subtractive interference by path identity (PSIPI) has traits of two distinct types of interference phenomena: 1) interference of fields generated by independent sources \cite{pfleegor1967interference,Mandel1983,ou1997parametric,chrapkiewicz2016hologram,ou2007multi} and 2) interference by path identity \cite{zou1991induced,wang1991induced,Lahiri2022PI_RMP}. Specifically, the phase-subtractive nature is due to the presence of two independent sources, and the fact that the interference pattern contains phase-information that was never introduced to the detected photons is due to path identity. We stress that path identity is absolutely essential to observe this interference effect. In fact, it can be checked using Eqs.~(\ref{2-mode-state}) and \eqref{coin-form} that the state given by Eq.~(\ref{2-mode-state}), which is obtained without path identity, does not lead to interference. 
\par
Let us also note that one arrives at the same expression for the interference pattern [Eq.~\eqref{2-ph-PI}] if, instead of Eq.~\eqref{2-mode-state-PI}, one employs the quantum state obtained by applying path identity [Eq.~\eqref{PI-state}] to Eq.~\eqref{full-2mode-state-supp}. In this case, one finds that the contribution from any term that appears in the latter state but not in Eq.~(\ref{2-mode-state-PI}) is identically zero. Notably, these non-contributing terms contain the states, $\ket{2S_u,2I_{u'}}_1$, $\ket{2S_v,2I_{v'}}_1$, $\ket{2S_c,2I_{c'}}_2$, and $\ket{2S_d,2I_{d'}}_2$, which are forbidden in a fermionic system. This fact indicates that the phenomenon of PSIPI can be observed with fermions. In fact, fermions can legitimately be in the quantum states given by Eqs.~\eqref{2-mode-state} and (\ref{2-mode-state-PI}). Furthermore, bosonic creation and annihilation operators used in Eqs.~\eqref{det-field} and \eqref{coin-form} can be replaced with fermionic creation and annihilation operators leading to the same expression for the probability of coincidence detection [Eq.~\eqref{2-ph-PI}]. We, therefore, conclude that PSIPI can also be observed with fermionic systems.

\section{Summary and Outlook}\label{summary}
We have reported a unique two-particle interference phenomenon that employs four-particle states generated by two independent sources and is enabled by path identity of two undetected particles. We have also introduced a noise-resistant quantitative phase imaging technique that  acquires images at wavelengths for which no detectors are available. Since quantitative phase imaging is used in biological and medical research \cite{Park2018}, for example, in immunology \cite{mitchell2018nongenetic}, cell biology \cite{mir2011optical}, and cancer diagnosis \cite{majeed2018label} and prognosis \cite{uttam2015early}, we expect that our findings will have applications in these fields. Our results are applicable to any modality where the principle of standard QIUP is used to retrieve object information, e.g., spectroscopy \cite{kalashnikov2016infrared}, microscopy \cite{kviatkovsky2020microscopy,paterova2020hyperspectral}, holography \cite{topfer2022quantum} and optical coherence tomography \cite{valles2018optical,paterova2018tunable}. Finally, interference by path identity has been applied to fundamental tests of quantum mechanics \cite{heuer2015induced,heuer2015complementarity}, and also to measurement \cite{pol-ent-theory,pol-ent-exp,rajeev2023single} and generation \cite{lahiri2018many,MarioPhysRevLett2017,kysela2020path,zhang2012way,Paraoanu_many-part-ent_PRA} of quantum entanglement. It will be interesting to explore the implications of our results for these topics.  	
\vskip 0.5cm
\section*{Acknowledgment} 
The research was supported by the U.S. Office of Naval Research under award number N00014-23-1-2778.

\appendix
\setcounter{figure}{0} 
\renewcommand{\thefigure}{\thesection\arabic{figure}}

\section{Derivation of Eq.~\eqref{2-ph-pattern}}\label{sec:StandardInterference}
In this appendix, we derive Eq.~\eqref{2-ph-pattern}, which represents a typical two-particle interference pattern. The basic features of a standard two-particle interferometer (Fig.~\ref{fig:2mode}a) have been described in Sec.~\ref{sec:two-part-interf} of the main text. When the two sources in the interferometer emit coherently and with equal probability, the two-particle state generated by them is given by \cite{horne1989two}
\begin{align}\label{tw-ph-state-supp} 
\ket{\psi}=\frac{1}{\sqrt{2}}(\ket{S_u}_1\ket{S_v}_1+\ket{S_c}_2\ket{S_d}_2),
\end{align}
where $\ket{S_u}_1$ denotes a single particle $S$ that is emitted from $Q_1$ and in path $u$, etc.
\par
The field operators (bosonic or fermionic) at the detectors placed at beamsplitter outputs $h$ and $g$ are, respectively, given by the following standard expressions: $\opa_{S}(h) \propto \opa_{S_1}(u)+ie^{i\phi_S}\opa_{S_2}(c)$ and $\opa_{S}(g) \propto \opa_{S_1}(v)+ie^{i\phi_S'}\opa_{S_2}(d)$. Here, $\opa_{S_1}(u)$ represents the annihilation operator for a $S$-particle emitted by source $Q_1$ in mode $u$, etc. 
\par
The probability of joint detection of two $S$-particles at $h$ and $g$ can now be obtained using the standard expression $P_{hg}\equiv \bra{\psi}\opa^\dagger_S(h)\opa^\dagger_S(g)\opa_S(g)\opa_S(h)\ket{\psi}$, where $\ket{\psi}$ is given by Eq.~(\ref{tw-ph-state-supp}). The resulting probability, which is proportional to the coincidence counting rate, is given by Eq.~\eqref{2-ph-pattern} in the main text.

\section{Recollection of some basic results from the theory of SPDC}\label{sec:SPDC}
In this section, we recollect relevant results form the theory of spontaneous parametric down-conversion (SPDC) in a nonlinear crystal. The scalar treatment is enough for our purpose. 
\par
In the interaction picture, the Hamiltonian of the SPDC process can be written as \cite{ghosh1986twophotonSPDC,walborn2010spatial}
\begin{align}\label{Hamiltonian}
\hat{H}_{\text{in}}(t) &= \sum_{\k_p,\k_S,\k_I}V_p(\k_p)f(\omega_S,\omega_I;\omega_p)\nonumber\\
&\times\exp[i(\omega_S+\omega_I-\omega_P)t]\ocr_S(\k_S)\ocr_I(\k_I)\nonumber\\
&\times\int_D\text{d}^3r~\exp[i(\k_P-\k_S-\k_I)\cdot\r)]+ \text{H.c.},
\end{align}
where the labels $p,S,I$ correspond to pump, signal, and idler photons, respectively; $V_p(\k_p)$ is the amplitude of the scalar pump field, $f(\omega_S,\omega_I;\omega_p)$ represents the interaction strength between the pump field and the nonlinear crystal, $\opa_S^\dagger(\k)$ ($\opa_I^\dagger(\k)$) creates a signal (idler) photon in mode $\k$, $D$ is the volume of the crystal, and H.c. represents Hermitian conjugate.
\par
The quantum state of the light generated by down conversion at the crystal is then given by
\begin{align}\label{state-int-supp}
\ket{\psi} = \hat{U}\ket{\text{vac}},
\end{align}
where $\ket{\text{vac}}$ is the vacuum state, and $\hat{U}(t)$ is given by the standard perturbative expression (see, for example, Ref. \cite{cohen-tannoudji})
\begin{align}\label{unitary-supp}
&\hat{U}= \hat{\openone}+\frac{1}{i\hbar}\int_0^\tau \text{d}t\hat{H}_{\text{in}}(t) \nonumber\\
&+\left(\frac{1}{i\hbar}\right)^2\int_0^\tau \text{d}t\int_0^t \text{d}t'\hat{H}_{\text{in}}(t)\hat{H}_{\text{in}}(t')+\cdots ,
\end{align}
where $\tau$ is the interaction time.
\par
Evaluating the integrals allows the operator to be expressed in the simpler form
\begin{align}\label{simple-uhat-supp}
    &\hat{U} = \hat{\openone}+\sum_{\k_S,\k_I}C^{(2)}(\k_S,\k_I)\opa^\dagger_{S}(\k_S)\opa^\dagger_{I}(\k_I)\nonumber\\
    &+\sum_{\substack{\k_S,\k_S'\\ \k_I,\k_I'}}C^{(4)}(\k_S,\k_S',\k_I,\k_I')\opa^\dagger_{S}(\k_S)\opa^\dagger_{S}(\k_S')\opa^\dagger_{I}(\k_I)\opa^\dagger_{I}(\k_I')\nonumber\\
    &+\text{H.c.}+\cdots,
\end{align}
where
\begin{subequations}\label{C-coefficients-supp}
	\begin{align}
	&C^{(2)}(\k_S,\k_I) = \frac{D\tau}{i\hbar}\sum_{\k_p} V_p(\k_p) f(\omega_S,\omega_I;\omega_p)\nonumber\\
	& \qquad \qquad \times\exp[i\Delta\omega \tau/2]\exp[-i\Delta\k\cdot\r_0]\nonumber\\
	& \qquad \qquad \times\sinc(\Delta\omega \tau/2)\prod_{m=1}^3\sinc(\Delta k_m\ell_m/2) \label{C-coefficients-supp:a}, \\
	&C^{(4)}(\k_S,\k_S',\k_I,\k_I') = \frac{1}{2}C^{(2)}(\k_S,\k_I)C^{(2)}(\k_S',\k_I').\label{C-coefficients-supp:b}
	\end{align}
\end{subequations}
In Eqs.~(\ref{C-coefficients-supp:a}) and (\ref{C-coefficients-supp:b}), $\Delta\omega\equiv\omega_S+\omega_I-\omega_p$, $\Delta\k\equiv\k_S+\k_I-\k_p$, and $\r_0$ represents the position of the center of the crystal with side lengths of $\ell_1$, $\ell_2$, and $\ell_3$. Note that $C^{(2)}$ can be modeled by well-behaved functions which agree very well with experimental observations \cite{walborn2010spatial}.
\par
This allows us to use Eq.~(\ref{state-int-supp}) to write the state as
\begin{align}\label{one-source-state-supp}
&\ket{\psi} \propto \ket{\text{vac}}+\sum_{\k_S,\k_I}C^{(2)}(\k_S,\k_I)\ket{\k_S,\k_I} \nonumber\\
&+\sum_{\substack{\k_S,\k_S'\\\k_I,\k_I'}}C^{(4)}(\k_S,\k_S',\k_I,\k_I')\ket{\k_S,\k_S',\k_I,\k_I'}+\cdots.
\end{align}
\par
\textbf{Perfect Momentum Correlation.} In the case of perfect momentum correlation, given a particular idler photon wave vector, $\k_I$, we know with certainty that the corresponding signal photon wave vector will have a fixed value of $\k_S$. Such a situation is attained when the the spatial phase-matching condition $\k_p=\k_S+\k_I$ holds to high accuracy and the pump beam is very well collimated such that the pump field can be practically assumed to contain only one wave vector $\k_p=\k_0$. In this case, the term $\prod_{m=1}^3 \sinc(\Delta k_m\ell_m/2)$ in Eq.~\eqref{C-coefficients-supp:a} can be treated as a delta function, that is, Eq.~\eqref{C-coefficients-supp:a} can be expressed as
\begin{align}\label{perf-MC}
C^{(2)}(\k_S,\k_I)= V_p \,g(\k_S,\k_I) \, \delta_{\k_S,\k_0-\k_I},
\end{align}
where $g(\k_S,\k_I)$ represents the rest of the terms on the right-hand side of Eq.~\eqref{C-coefficients-supp:a}, and $V_p(\k_0)\equiv V_p$. Note that $g(\k_S,\k_I)$ is always proportional to the gain of the crystal.
\par
If we now substitute for $C^{(2)}(\k_S,\k_I)$ from Eq.~\eqref{perf-MC} into Eq.~\eqref{one-source-state-supp} and use Eq.~\eqref{C-coefficients-supp:b}, we obtain
\begin{align}\label{multimode-perf-corr}
    &\ket{\psi'} = \ket{\text{vac}} + V_p\sum_{\k_I}g(\k_S,\k_I)\ket{\k_S,\k_I}\nonumber\\
    &+\frac{V_p^2}{2}\sum_{\k_I,\k_I'}g(\k_S,\k_I)g(\k_S',\k_I')\ket{\k_S,\k_S',\k_I,\k_I'}+\cdots,
\end{align}
where $\k_S=\k_0-\k_I$ and $\k_S'=\k_0-\k_I'$.

\section{SPDC in perfectly correlated two-mode scenario}\label{sec:SPDC-2mode}
In this Appendix, we consider the case of perfect momentum correlation in a two-mode scenario and show how Eq.~\eqref{2mode-state-1} is obtained from Eq.~\eqref{multimode-perf-corr}. 
\par
We suppose that we have only two possible modes into which each of the signal and idler photons can emit from a given source. For photons generated at source $Q_1$ ($Q_2$), we label the signal modes $u,v$ ($c,d$) and the corresponding idler modes $u',v'$ ($c',d'$) as in the main text. The perfect correlation condition in this notation is taken to mean that a signal photon in path $u$ is always accompanied by an idler photon in mode $u'$ and so on. In principle, this leads to different values of the function $g(\k_S,\k_I)$ defined above, namely, $g(u,u')$, $g(v,v')$, $g(c,c')$, and $g(d,d')$. To simplify discussion in this two-mode scenario, we make the assumption that each of these values is approximately equal, i.e., $g(u,u')\approx g(v,v')\approx \cdots\equiv g$. 
\par
Under this additional assumption, the perfectly-correlated multimode state, Eq.~\eqref{multimode-perf-corr}, produced at source $Q_1$ takes the form
\begin{align}\label{state-Q1-supp}
    \ket{\psi''}_1 &=\ket{\text{vac}} + gV_p \left(\ket{S_u,I_{u'}}_1+\ket{S_v,I_{v'}}_1 \right)\nonumber\\
    &+g^2V_p^2\Big(\ket{2S_u,2I_{u'}}_1+ \ket{2S_v,2I_{v'}}_1 \nonumber \\ & \qquad \quad \quad + \ket{S_u,I_{u'},S_v,I_{v'}}_1 \Big)+\cdots, 
\end{align}
where we have expanded the summation, and we recall that $[\opa^\dagger]^2\ket{\text{vac}}=\sqrt{2}\ket{2}$. This is equivalent to Eq.~\eqref{2mode-state-1} of the main text.
\par
For completeness, by following a similar procedure to the above, the state produced by photons generated at source $Q_2$ is given by
\begin{align}\label{state-Q2-supp}
    \ket{\psi''}_2 &=\ket{\text{vac}} + gV_pe^{i\widetilde{\Theta}} \left(\ket{S_c,I_{c'}}_1+\ket{S_d,I_{d'}}_1 \right)\nonumber\\
    &+g^2V_p^2e^{2i\widetilde{\Theta}}\Big(\ket{2S_c,2I_{c'}}_1+ \ket{2S_d,2I_{d'}}_1 \nonumber \\ & \qquad \quad \quad + \ket{S_c,I_{c'},S_d,I_{d'}}_1 \Big)+\cdots,
\end{align}
where we recall that the pump field at source $Q_2$ differs from that at $Q_1$ by a randomly fluctuating phase, $\widetilde{\Theta}$, such that $V_{p_2}=V_pe^{i\widetilde{\Theta}}$.

\section{Further details on coincidence counting rate in the multimode scenario}\label{coin-mm-app}
We now calculate the coincidence count rates corresponding to detections at the different possible pairs of outputs ($b,b'$) of the beamsplitter in the setup shown by Fig.~\ref{fig:RandPhase}b. A general expression for the coincidence counting rate at a pair of points $(\r_{\k_S},\r_{\k_S'})$ is given by 
\begin{align}\label{coin-2-app}
&P_{jl}^{(2)}(\r_{\k_S},\r_{\k_S'}) \nonumber\\
&=\bra{\psi^{(4)}}\opEns_l(\r_{\k_S'})\opEns_j(\r_{\k_S})\opEps_j(\r_{\k_S'})\opEps_l(\r_{\k_S})\ket{\psi^{(4)}},
\end{align}
where $j=b,b'$ and $l=b,b'$ represent outputs of the beamsplitter, $\opEps_j(\r_{\k_S'})$ represents the positive frequency part of the quantized electric field at output $j$, $\opEns=\{\opEps\}^\dagger$, and $\ket{\psi^{(4)}}$ is given by Eq.~(\ref{state-four-ZWM-Supp}) of the main text. We note that $P_{jl}^{(2)}(\r_{\k_S},\r_{\k_S'})=P_{lj}^{(2)}(\r_{\k_S},\r_{\k_S'})$. The field operators at outputs $b$ and $b'$ can be written as
\begin{subequations}\label{fields-app}
\begin{align}
\opEps_b(\r_{\k_S})\propto \opa_{S_1}(\k_S)+ie^{i\phi_S}\opa_{S_2}(\k_S),\label{b_output-supp} \\
\opEps_{b'}(\r_{\k_S})\propto i\opa_{S_1}(\k_S)+e^{i\phi_S}\opa_{S_2}(\k_S).\label{bp_output-supp}
\end{align}
\end{subequations}
\par
We first consider the case in which both points $(\r_{\k_S},\ \r_{\k_S'})$ are located at the output port $b$. Using Eqs.~(\ref{state-four-ZWM-Supp}),~(\ref{coin-2-app}), and (\ref{fields-app}), the coincidence counting rate is given by 
\begin{align}\label{coin-rate-app}
    &P_{bb}^{(2)}(\r_{\k_S},\r_{\k_S'}) \propto \sum_{\k_I,\k_I'}\bigg[P^{(4)}_{1}(\k_S,\k_S',\k_I,\k_I')\nonumber\\
    &+P^{(4)}_{2}(\k_S,\k_S',\k_I,\k_I') \nonumber\\
    &+2P^{(4)}_{12}(\k_S,\k_S',\k_I,\k_I') \big\{1+\cos [\alpha(\r_{\k_I})-\alpha(\r_{\k_I'})]\big\}\bigg],
\end{align}
where $P^{(4)}_{1}$ and $P^{(4)}_{2}$ are contributions from NL1 and NL2, respectively, and the term  containing $P^{(4)}_{12}$ is due to joint emissions at NL1 and NL2. Equation (\ref{coin-rate-app}) is equivalent to Eq.~\eqref{coin-rate} of the main text. The explicit forms of $P^{(4)}_{1}$, $P^{(4)}_{2}$, and $P^{(4)}_{12}$ are given by
\begin{subequations}\label{Pj}    
\begin{align}       
&P^{(4)}_1(\k_S,\k_S',\k_I,\k_I') = \big\vert C^{(4)}(\k_S,\k_S',\k_I,\k_I')\nonumber\\
&+C^{(4)}(\k_S,\k_S',\k_I',\k_I)+C^{(4)}(\k_S',\k_S,\k_I,\k_I')\nonumber\\
&+C^{(4)}(\k_S',\k_S,\k_I',\k_I)\big\vert^2, \label{Pj:a} 
\end{align}
\begin{align}
&P^{(4)}_2(\k_S,\k_S',\k_I,\k_I') = \big\vert C^{(4)}(\k_S,\k_S',\k_I,\k_I')\nonumber\\
&+C^{(4)}(\k_S,\k_S',\k_I',\k_I)+C^{(4)}(\k_S',\k_S,\k_I,\k_I')\nonumber\\
&+C^{(4)}(\k_S',\k_S,\k_I',\k_I)\big\vert^2, \label{Pj:b} 
\end{align}
\begin{align}
&P_{12}^{(4)}(\k_{S},\k_{S}',\k_{I},\k_{I}') = \big\vert C^{(2)}(\k_S,\k_I)C^{(2)}(\k_S',\k_I') \nonumber\\
&+C^{(2)}(\k_S,\k_I')C^{(2)}(\k_S',\k_I)\big\vert^2. \label{Pj:c}
\end{align}
\end{subequations}
We now show the expressions for the other possible coincidence count rates.

\par
We next consider the case in which both points are located at the output $b'$. From Eqs.~(\ref{state-four-ZWM-Supp}), (\ref{coin-2-app}), and (\ref{bp_output-supp}), it follows that 
\begin{align}\label{P-alt-supp}
&P_{b'b'}^{(2)}(\r_{\k_S},\r_{\k_S'}) \propto \sum_{\k_I,\k_I'}\bigg[P^{(4)}_{1}(\k_S,\k_S',\k_I,\k_I')\nonumber\\
    &+P^{(4)}_{2}(\k_S,\k_S',\k_I,\k_I') \nonumber\\
    &+2P^{(4)}_{12}(\k_S,\k_S',\k_I,\k_I') \big\{1+\cos [\alpha(\r_{\k_I})-\alpha(\r_{\k_I'})]\big\}\bigg].
\end{align}
It is to be noted that $P^{(2)}_{b'b'}$ has the same expression as $P^{(2)}_{bb}$.
\par
We finally determine the coincidence counting rate for the case in which one point is located at $b$ and the other at $b'$. Using Eqs.~(\ref{state-four-ZWM-Supp}), (\ref{coin-2-app}), and (\ref{fields-app}), we find that
\begin{align}
&P_{bb'}^{(2)}(\r_{\k_S},\r_{\k_S'}) \propto \sum_{\k_I,\k_I'}\bigg[P^{(4)}_{1}(\k_S,\k_S',\k_I,\k_I')\nonumber\\
    &+P^{(4)}_{2}(\k_S,\k_S',\k_I,\k_I') \nonumber\\
    &+2\widetilde{P}^{(4)}_{12}(\k_S,\k_S',\k_I,\k_I') \big\{1-\cos [\alpha(\r_{\k_I})-\alpha(\r_{\k_I'})]\big\}\bigg],
\end{align}
where 
\begin{align}
\widetilde{P}^{(4)}_{12}(\k_S,\k_S',\k_I,\k_I')&=\big\vert C^{(2)}(\k_S,\k_I)C^{(2)}(\k_S',\k_I')\nonumber\\
&-C^{(2)}(\k_S,\k_I')C^{(2)}(\k_S',\k_I)\big\vert^2.
\end{align}

\section{Density matrix treatment}\label{sec:DMT}
In the main text, we performed the analysis by introducing a stochastic phase $\widetilde{\Theta}$ and applying the existing formalism involving pure states. This treatment is equivalent to a more formal approach that involves mixed states. In this section, we outline the mixed-state-based approach.
\par
Suppose that three states $\ket{\psi_1},~\ket{\psi_2},$ and $\ket{\psi_3}$ are emitted with probability amplitudes $\alpha_1,~\alpha_2$, and $\alpha_3$, respectively. If the emissions are independent (i.e., mutually incoherent), the quantum state can be expressed as
\begin{align}\label{basic-mixed-state}
    \hat{\rho}=\vert\alpha_1\vert^2\ket{\psi_1}\bra{\psi_1}+\vert\alpha_2\vert^2\ket{\psi_2}\bra{\psi_2}+\vert\alpha_3\vert^2\ket{\psi_3}\bra{\psi_3}.
\end{align}
\par
In our case (Fig.~\ref{fig:2mode}b), there are three such states that contribute to coincidence counts at $g$ and $h$ (see also the discussion under Sec.~\ref{subsec:state-no-PI} above): (1) four-photons emitted at $Q_1$, creating $\ket{\psi_1}=\ket{S_u,S_v,I_{u'},I_{v'}}_1$, (2) four-photons emitted at $Q_2$, creating $\ket{\psi_2}=\ket{S_c,S_d,I_{c'},I_{d'}}_2$, and (3) simultaneous single-pair productions at $Q_1$ and $Q_2$ with two options, creating $\ket{\psi_3}=\ket{S_u,I_{u'}}_1\ket{S_d,I_{d'}}_2+\ket{S_v,I_{v'}}_1\ket{S_c,I_{c'}}_2$. Since these three emissions are independent of each other, we can apply Eq.~(\ref{basic-mixed-state}) and obtain the resulting density operator as
\begin{align}\label{2mode-density-matrix} \hat{\rho}'\propto&\ket{S_u,S_v,I_{u'},I_{v'}}_{1}\bra{S_u,S_v,I_{u'},I_{v'}}_1\nonumber\\
    &+\ket{S_c,S_d,I_{c'},I_{d'}}_{2}\bra{S_c,S_d,I_{c'},I_{d'}}_2\nonumber\\
    &+\bigg\{\ket{S_u,I_{u'}}_1\ket{S_d,I_{d'}}_{2}\bra{S_d,I_{d'}}_2\bra{S_u,I_{u'}}_1\nonumber\\
    &+\ket{S_v,I_{v'}}_1\ket{S_c,I_{c'}}_{2}\bra{S_c,I_{c'}}_2\bra{S_v,I_{v'}}_1\nonumber\\
    &+\big[\ket{S_u,I_{u'}}_1\ket{S_d,I_{d'}}_{2}\bra{S_c,I_{c'}}_2\bra{S_v,I_{v'}}_1+\text{H.c.}\big]\bigg\},
\end{align}
where we have dropped a normalization coefficient and have assumed that the three states are emitted with equal probability. 
\par
We now apply the relation between the kets due to path identity [Eq.~\eqref{PI-state} in main text]. Using Eqs.~\eqref{PI-state} and (\ref{2mode-density-matrix}), we find the quantum state of the system takes the form
\begin{align}\label{path-aligned-2mode-dm}
    \hat{\rho}\propto&\ket{I_{u'},I_{v'}}_{1}\bra{I_{u'},I_{v'}}_1\otimes\bigg\{\ket{S_u,S_v}_{1}\bra{S_u,S_v}_1\nonumber\\
    &+\ket{S_c,S_d}_{2}\bra{S_c,S_d}_2+\ket{S_u}_1\ket{S_d}_{2}\bra{S_d}_2\bra{S_u}_1\nonumber\\
    &+\ket{S_v}_1\ket{S_c}_{2}\bra{S_c}_2\bra{S_v}_1\nonumber\\
    &+\big[e^{i(\gamma_I-\gamma_I')}\ket{S_u}_1\ket{S_d}_{2}\bra{S_c}_2\bra{S_v}_1+\text{H.c.}\big]\bigg\},
\end{align}
where we have dropped a normalization coefficient.
\par
The coincidence counting rate at the pair of outputs $h$ and $g$ is given by
\begin{align}\label{2mode-coin-rate-trace}  P_{hg}\propto\text{tr}\big\{\hat{\rho}\thinspace \opa^\dagger_S(h)\opa^\dagger_S(g)\opa_S(g)\opa_S(h)\big\},
\end{align}
where $\opa_S(g)$ and $\opa_S(h)$ are defined in Eqs.~(\ref{det-field:a}) and (\ref{det-field:b}). Equations (\ref{path-aligned-2mode-dm}) and (\ref{2mode-coin-rate-trace}) give Eq.~\eqref{2-ph-PI} of the main text, i.e., 
\begin{align}
    P_{hg}\propto 1+\frac{1}{2}\cos(\phi_S'-\phi_S+\gamma_I-\gamma_I').
\end{align}
\par
If we follow the same procedure for the multi-mode photonic state derived in Eq.~\eqref{state-four-ZWM-Supp}, we find that
\begin{align}\label{density-matrix-four-Supp}
    &\hat{\rho} \propto \nonumber\\
    &\sum_{\substack{\k_S,\k_S' \\ \k_I,\k_I'}}\sum_{\substack{\k_S'',\k_S''' \\ \k_I'', \k_I'''}}\bigg\{[C^{(4)}(\k_S'',\k_S''',\k_I'',\k_I''')]^*C^{(4)}(\k_S,\k_S',\k_I,\k_I')\nonumber\\
    &\qquad \times\big(\ket{\k_S,\k_S'}_{1}\bra{\k_S'',\k_S'''}_1\nonumber\\
    &\qquad+e^{i[\alpha(\r_{\k_I''})+\alpha(\r_{k_I'''})-\alpha(\r_{k_I})-\alpha(\r_{k_I'})]}\ket{\k_S,\k_S'}_{2}\bra{\k_S'',\k_S'''}_2\big) \nonumber\\
    &+[C^{(2)}(\k_S'',\k_I'')C^{(2)}(\k_S''',\k_I''')]^*C^{(2)}(\k_S,\k_I)C^{(2)}(\k_S',\k_I') \nonumber\\
    &\qquad\times e^{i[\alpha(\r_{k_I'''})-\alpha(\r_{k_I'})]}\ket{\k_S}_{1}\bra{\k_S''}_1\otimes\ket{\k_S'}_{2}\bra{\k_S'''}_2\bigg\}\nonumber\\
    &\otimes\ket{\k_I,\k_I'}_{1}\bra{\k_I'',\k_I'''}_1.
\end{align}
\par
In this case, the coincidence counting rate at a pair of points $(\r_{\k_S},\r_{\k_S'})$ is given by
\begin{align}\label{multimode-coin-rate-dm}
    P^{(2)}_{\ell m}(\r_{\k_S},\r_{\k_S'})\propto\text{tr}\big\{&\hat{\rho}\opEns_\ell(\r_{\k_S})\opEns_m(\r_{\k_S'})\nonumber\\
    &\times\opEps_m(\r_{\k_S'})\opEps_\ell(\r_{\k_S})\big\},
\end{align}
where $\ell=b,b';~m=b,b'$. If we set $\ell=m=b$, it follows from Eqs. (\ref{density-matrix-four-Supp}), (\ref{multimode-coin-rate-dm}), and \eqref{fields-app} that
\begin{align}\label{coin-rate-dm}
	&P_{bb}^{(2)}(\r_{\k_S},\r_{\k_S'}) \propto \sum_{\k_I,\k_I'}\bigg[P^{(4)}_{1}(\k_S,\k_S',\k_I,\k_I')\nonumber\\
    &+P^{(4)}_{2}(\k_S,\k_S',\k_I,\k_I') \nonumber\\
    &+2P^{(4)}_{12}(\k_S,\k_S',\k_I,\k_I') \big\{1+\cos [\alpha(\r_{\k_I})-\alpha(\r_{\k_I'})]\big\}\bigg],
\end{align}
which is the same expression as Eq.~\eqref{coin-rate} in the main text.

\vskip 0.5cm
\section{An alternative setup to realize phase-subtractive two-photon interference by path identity and noise-resistant quantum phase imaging with undetected photons}\label{sec:HRWZ}
\begin{figure}[htbp]
	\includegraphics[width=\linewidth]{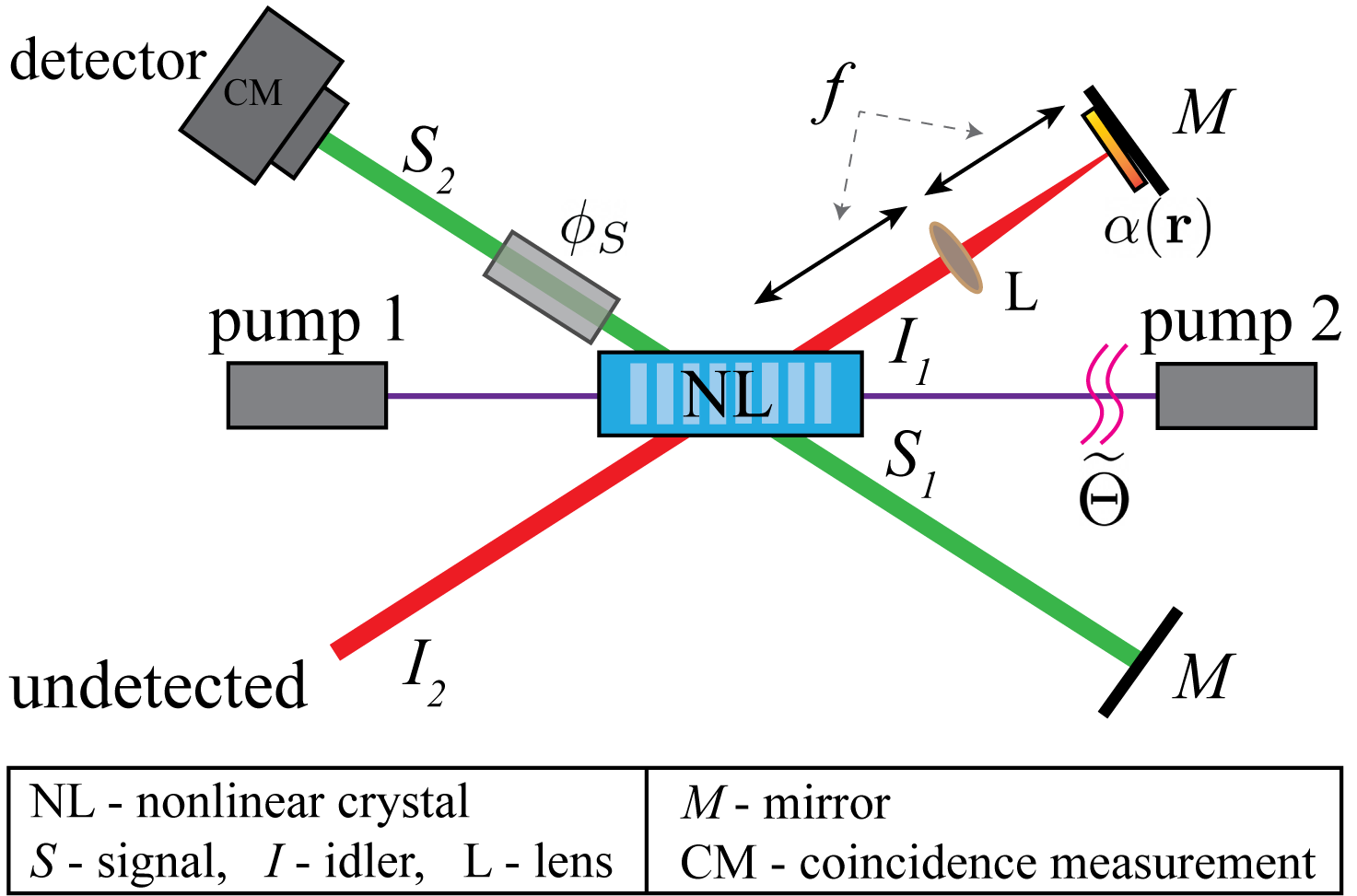}
	\caption{An alternative setup that employs a single crystal. The crystal is pumped from two sides by two mutually-incoherent pump beams with stochastic phase difference $\widetilde{\Theta}$. Mirrors are used to align paths of signal and idler photons by reflecting them back through the crystals. Signal photons are then detected, while idler photons are not.}
	\label{fig:HRWZ-supp}
\end{figure}

In this section, we present an alternative setup (Fig.~\ref{fig:HRWZ-supp}) that can be used to experimentally realize phase-subtractive interference by path identity and noise-resistant phase imaging with undetected photons. The setup is inspired by an experiment performed by Herzog et al. \cite{herzog1994frustrated}. 
\par
In contrast to the setup presented by Fig.~\ref{fig:RandPhase}b of the main text, this setup (Fig.~\ref{fig:HRWZ-supp}) employs only one nonlinear crystal. Furthermore, the geometry of the setup discussed in the main text is similar to that of a Mach-Zehnder interferometer; whereas, the geometry of the setup presented here is more similar to that of a Michelson interferometer. Despite these differences, these two setups work under the same principle. 
\par
As shown in Fig.~\ref{fig:HRWZ-supp}, a nonlinear crystal is pumped from two sides by two \emph{mutually incoherent} pump beams. We call these two pump beams pump 1 and pump 2. Signal and idler photons generated by SPDC due to pump 1 propagate in beams $S_1$ and $I_1$, respectively. Likewise, signal and idler photons  generated by SPDC due to pump 2 propagate in beams $S_2$ and $I_2$, respectively. The four-photon quantum state generated by these two SPDC processes before path identity is applied is identical to that obtained for the setup given in the main text. That is, the state is given by Eq.~(\ref{state-four-ZWM-Supp}). 
\par
Idler beam $I_1$ is reflected by a mirror and sent back through the crystal in such a way that it perfectly overlaps with beam $I_2$ (Fig.~\ref{fig:HRWZ-supp}). Signal beams are also overlapped following the same procedure. We thus have two path identity relations
\begin{subequations}\label{HRWZ-alignment-supp}
	\begin{align}
	&\opa_{I_2}(\k_I) = e^{2i\alpha(\r_{\k_I})}\opa_{I_1}(\k_I)\label{HRWZ-alignment-supp:a}, \\
	&\opa_{S_2}(\k_S) = e^{i\phi_S}\opa_{S_1}(\k_S),\label{HRWZ-alignment-supp:b}
	\end{align}
\end{subequations}
where $\phi_S$ is the phase acquired by signal photons, which can be made to be spatially-independent with highly-collimated beams, and $\alpha$ is the spatially-dependent phase. Note that a factor of two comes in front of $\alpha$ because the mirror sends the idler beam through the object twice.
\par
The resulting analysis is similar to that given in the main text; the only difference is in the path identity relation. We find that the coincidence counting rate at two points $(\r_{\k_S},\r_{\k_S'})$ on the detector is given by
\begin{align}\label{coin-rate-HRWZ-supp}
&P^{(2)}(\r_{\k_S},\r_{\k_S'}) \propto \sum_{\k_I,\k_I'}\bigg[P^{(4)}_{1}(\k_S,\k_S',\k_I,\k_I')\nonumber\\
    &+P^{(4)}_{2}(\k_S,\k_S',\k_I,\k_I') \nonumber\\
    &+2P^{(4)}_{12}(\k_S,\k_S',\k_I,\k_I') \big\{1+\cos 2[\alpha(\r_{\k_I})-\alpha(\r_{\k_I'})]\big\}\bigg],
\end{align}
where $P_1^{(4)},\ P_2^{(4)}$, and $P_{12}^{(4)}$ are given by Eqs.~(\ref{Pj:a})-(\ref{Pj:c}). If we compare Eq.~(\ref{coin-rate-HRWZ-supp}) with Eq.~\eqref{coin-rate} of the main text, we find that they are identical with the exception that $\alpha$ is replaced by $2\alpha$.

\bibliography{bib-arXiv-3.bib}
	
\end{document}